\documentclass[conference]{IEEEtran}
\usepackage{multirow}
\usepackage{graphicx}
\usepackage{caption}
\usepackage{subcaption}
\usepackage{tabularx}

\begin{document}

\title{Is There WiFi Yet? How Aggressive WiFi Probe Requests Deteriorate Energy and Throughput}

%
%
%
%
%

\author{\IEEEauthorblockN{Xueheng Hu, Lixing Song, Dirk Van Bruggen, Aaron Striegel}
\IEEEauthorblockA{Department of Computer Science and Engineering\\
University of Notre Dame\\
Notre Dame, IN USA\\
Email: xhu2@nd.edu, lsong2@nd.edu, dvanbrug@alumni.nd.edu, striegel@nd.edu}
}



\maketitle

\begin{abstract}
WiFi offloading has emerged as a key component of cellular operator strategy to meet the data needs of rich, mobile devices. As such, mobile devices tend to aggressively seek out WiFi in order to provide improved user Quality of Experience (QoE) and cellular capacity relief. For home and work environments, aggressive WiFi scans can significantly improve the speed on which mobile nodes join the WiFi network. Unfortunately, the same aggressive behavior that excels in the home environment incurs considerable side effects across crowded wireless environments. In this paper, we show through empirical studies at both large (stadium) and small (laboratory) scales how aggressive WiFi scans can have significant implications for energy and throughput, both for the mobile nodes scanning and other nearby mobile nodes. We close with several thoughts on the disjoint incentives for properly balancing WiFi discovery speed and ultra-dense network interactions.        

\end{abstract}


\section{Introduction}

The past several years have seen a veritable explosion of data consumption on mobile devices.  Smartphones, tablets, and more recently the Internet of Things (IoT) have created a nearly insatiable demand for ubiquitous wireless connectivity.   While peak speeds for cellular (LTE) have risen impressively in the past few years, dense and indoor environments remain challenging scenarios for carriers. Although LTE-Advanced (LTE-A) will offer relief with the introduction of small cell support, questions remain with regards to small cell economic viability and management complexity \cite{Andrews:ChallengeCell}.

For the dense environment, WiFi offloading has emerged as a cornerstone of carrier strategy.  Despite the unlicensed nature of WiFi and potential issues with unpredictable QoE, the peak speeds of WiFi and more importantly the offloading of demand remain irresistible.  Hence, nearly all mobile devices aggressively push the user onto WiFi networks.  Whether it is prompting the user any time new WiFi is available, restricting certain services to WiFi only, the desire to offload is quite clear.  Further efforts by standards bodies such as ANDSF (Access Network Discovery and Service Function) \cite{ANDSF}, Hotspot 2.0 \cite{HS20}, and 802.11hew (High Efficiency Wireless) only reinforce that notion.

However, unlike cellular service, WiFi is neither pervasive nor contiguous.  Although ANDSF can effectively steer the user to WiFi is likely to be and Hotspot 2.0 can streamline the user joining, the mobile node must still find in the WiFi spectrum where the actual WiFi is located.  The root of this discovery process can be found in the basic 802.11 Probe Request whereby a mobile node will actively scan across the WiFi space (2.4, 5 GHz) for viable 802.11 access points (APs).  Access points, if inclined, can respond with Beacon Responses allowing the mobile device to quickly join the WiFi network rather than passively waiting to discover an AP.  Through the aggressive employment of active scans, mobile nodes can be rapidly offloaded to WiFi, satisfying both user QoE (Quality of Experience) and decreasing cellular network load.  For the ideal case of the home and work environment where the number of mobile devices is relatively low and the SSIDs are well known, such a setup tends to work fairly well.

Unfortunately, the tuning that is wonderful for the home and workplace tends to fare quite badly in the ultra-dense environment, despite the fact that the ultra-dense environment is where WiFi offloading is needed most.  In our paper, we show that not only do most mobile nodes excessively waste energy trying to find WiFi (is there WiFi yet) but those aggressive scans have significant secondary effects on the legitimate users of any established WiFi networks.  In short, our paper thesis is to argue that aggressive Probe Requests in the ultra-dense cases (hundreds or thousands of nodes) are the wireless equivalent of \emph{`Are we there yet?'}, just as annoying, wasteful, and infuriating but with significant implications for overall network health and performance.  To that end, the contributions of this paper are three-fold:

\begin{itemize}

\item \emph{Ultra-dense probe request dynamics:} We capture and analyze the dynamics of Probe Requests via packet captures at the entrance gates for two home football games.  We show that most mobile devices continue to unashamedly probe (intervals between scans of 20s or less) leading to considerable wasted energy and negative throughput implications.

\item \emph{Energy impact of probe requests:} We characterize the energy cost of active WiFi scanning, exploring the energy cost of a complete WiFi scan (Probe Request across all channels with appropriate SSIDs).  We show that aggressive scanning can burn up to 44\% more energy with little to no adaptation in response to the success or failure of WiFi scans.

\item \emph{Throughput impact of probe requests:} We isolate the negative effects of aggressive WiFi scanning across the 5 GHz band on network throughput.  We show that even with only a few mobile nodes under default settings can dramatically reduce network throughput.

\end{itemize}

\section {Related Work}
From an overall perspective, WiFi has received incredible attention from the research community. For the purposes of this paper though, we are chiefly concerned with works focusing on increased discovery speed \cite{Teng:INFOCOM09:DScan} and most notably, improved efficiency or accuracy for WiFi scanning \cite{Yeo:WiSe04:LANMonitor, Gupta:SECON07:WifiPhone, Raghavendra:TMC10:UnwantedLink, Rayanchu:NSDI12:WiFiNet,Cisco:Athens:80211ax}.  Our work is unique in that we highlight the prevalence of Probe Requests in the ultra-dense venue (affirmed by \cite{Cisco:Athens:80211ax}) as well as diving into the core impacts on energy and throughput.

The ability to efficiently and quickly scan is a fundamental requirement for fast, seamless handoffs in WiFi.  Teng. et. al in \cite{Teng:INFOCOM09:DScan} proposed D-Scan, specifically targeted at improving scan efficiency in dense environments. Monitoring also plays a key role in distinguishing performance issues with Yeo in \cite{Yeo:WiSe04:LANMonitor} and more contemporary work by Rayanchu et. al in \cite{Rayanchu:NSDI12:WiFiNet} trying to pin down interference issues creating issues with WiFi.  Two works from the literature are particularly relevant for this work. The work by Gupta and Mohapatra in \cite{Gupta:SECON07:WifiPhone} focused specifically on the power consumption of WiFi on phones while the work by \cite{Raghavendra:TMC10:UnwantedLink} looked at larger scale venues (IETF 2006 meeting) and overall performance.  The issue of ultra-dense venues and WiFi performance was recently discussed in a Cisco slide deck for the 802.11ax working group meeting in Athens in late 2014 \cite{Cisco:Athens:80211ax}.

\section{Ultra-Dense Dataset}

In this section, we analyze the data collected from two football games at the University of Notre Dame (Michigan on 09-06-14 and Stanford on 10-04-14). We begin with a general description of how the data was gathered and continue with in-depth analyses of the data.  

\subsection{Data Summary}

Data was gathered near gates to the football stadium using multiple Linux laptops with extended wireless adapters placed into monitor mode and running \emph{tcpdump}.  Notably, the stadium itself does not have publicly accessible WiFi hence offering a true picture of maximum number of Probe Requests in ultra-dense environments. The stadium itself seats roughly 80,000 with five separate entrance gates (A-E). Data gathering commences roughly one hour before the start of the game when most ticket holders begin to arrive at the stadium. Multiple laptops were used with Ubuntu 14.04 installed and each laptop possessed multiple external wireless adapters (TP-Link TL-WN7222N, Airpcap NX-900). Individual laptops were configured to monitor multiple bands, i.e., either monitor multiple 2.4 GHz channels (Channel 1, Channel 11) or across multiple bands (2.4 GHz Channel 1, 5 GHz Channel 153). Data is processed through a combination of \emph{tshark} and Python with Probe Request information stored in a MySQL database. Following processing, data files are discarded and only anonymized header information is preserved for the purpose of analysis.

\begin{table}
\centering
\caption{Ultra-Dense Data Summary}
\label{data_summary}

\begin{tabular}{|p{1.8cm}|p{0.001cm}|c|c|c|c|}
\hline

        \multicolumn{2}{|c|}{\textbf{Game}}                                              & \multicolumn{2}{c|}{Michigan}                                                   & \multicolumn{2}{c|}{Stanford}                                                   \\ \hline

      \multicolumn{2}{|c|}{\textbf{Time Duration}}                                  & \multicolumn{2}{c|}{28 min } &  \multicolumn{2}{c|}{42 min}                                                           \\ \hline

\multicolumn{2}{|c|}{\multirow{2}{*}{\textbf{Total \# of PRs}}} & 2.4G & 5G & 2.4G & 5G \\\cline{3-6} 

\multicolumn{2}{|c|}{}&78,175   & 14,441       & 86,195           & 4,335     \\ \hline

  \multicolumn{2}{|c|}{\textbf{Total \# of UEs} }    & 4,863& 2,716 & 6,813 &805\\\hline

\multirow{2}{*}{\textbf{PRs / Min}}                  & \multicolumn{1}{p{0.7cm}|}{Mean}                &    2,678           & 307                  & 2,168      & 97            \\ \cline{2-6} 

                                                     & \multicolumn{1}{p{0.7cm}|}{ Max}                &   3,703             & 596                & 3,029         & 223               \\ \hline
\end{tabular}
\end{table}

As shown in Table \ref{data_summary}, it is interesting to note the average number of Probe Requests across the 2.4 GHz and 5 GHz channels. Each number in the table represents the observations recorded for one particular channel (Channel 1 for 2.4 GHz, Channel 153 for 5 GHz). For the 2.4 GHz channel, the average density of Probe Requests comes in at 2678 per minute, just over 44 Probe Requests per second. The 5 GHz channel sees remarkably fewer Probe Requests (307 per minute) but it is also notable that many devices are still not fully 5 GHz capable. While we had expected to see a bump in 5 GHz Probe Requests at the Stanford game due to the recent release of iPhone 6, the inclement weather had clear impacts in terms of attendance for the game (upper 30s, rainy). Even with the reduced fan turnout, the number of Probe Requests for the Stanford game still averaged 2168 Probe Requests per minute in Channel 1 (nearly 36 Probe Requests per second).    

From an overhead perspective, each Probe Request can be viewed as entirely wasteful as no public WiFi exists at the stadium. From a distributional analysis (not shown due to space constraints), the most common rate setting (92\%) for Probe Requests in the 2.4 GHz spectrum was 1.0 Mb/s with speeds observed for Probe Requests up to 11.0 Mb/s. Probe Requests in the 5 GHz spectrum were universally set to 6.0 Mb/s. If we assume a rough Probe Request size of 100 bytes, a perfect Probe Request (ignoring CIFS, DIFS, 802.11 headers, DCF effects) would be 800 microseconds. Even with these largely unrealistic assumptions, the WiFi Probe Request overhead would be 3.5\% (35.2 milliseconds of Probe Request airtime per second). The reality though is that the impacts are much, much greater than the ideal 3.5\%.    

First, frequent Probe Requests are highly likely to impact the DCF of any mobile nodes affiliated with WiFi. While the stadium did not offer WiFi, we could view the mobile nodes as captured as being indicative of nodes without ANDSF policies / uncooperative mobile nodes.  Second, for the 2.4 GHz channels, the lack of channel orthogonality means that as a mobile node iterates through an active scan, it may cause issues as it traverses nearby channels (ex. Channel 2, Channel 3, Channel 4, Channel 5 on Channel 1).  Third, while the Probe Requests are relatively short, the low rates of the Probe Requests means that the actively scanning nodes tend to clutter / slow down the higher speed / affiliated nodes (ex. 1 Mb/s vs. 54 Mb/s).  Fourth, mobile nodes may continue to scan even once affiliated with an AP if AP performance is insufficient or simply if the mobile hopes for observing `better' WiFi.  As we will show later in the paper, the effects are quite significant with regards to throughput.

\subsection{Time Series Variation}

\begin{figure}

\centering

\begin{subfigure}[b]{0.45\textwidth}

\includegraphics[scale=0.65]{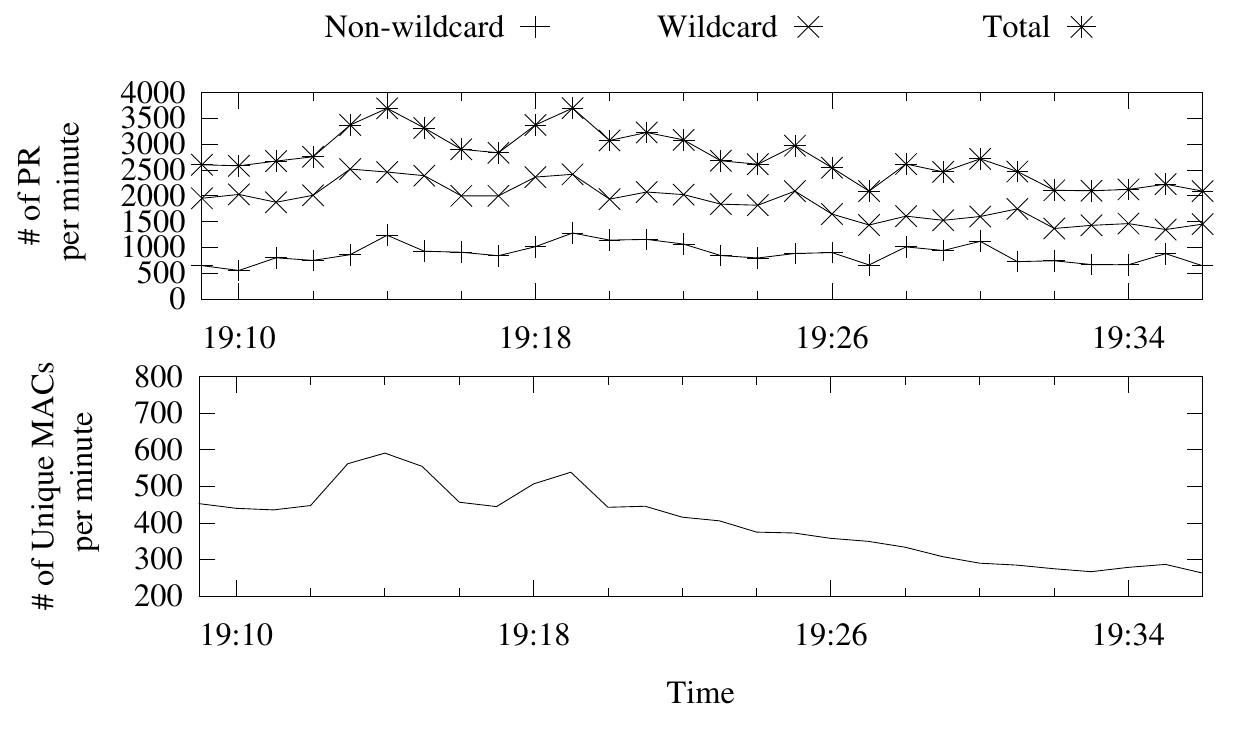}

\caption{Michigan game}
\label{Graph:MichiganTS}
\end{subfigure}

\begin{subfigure}[b]{0.45\textwidth}

\includegraphics[scale=0.65]{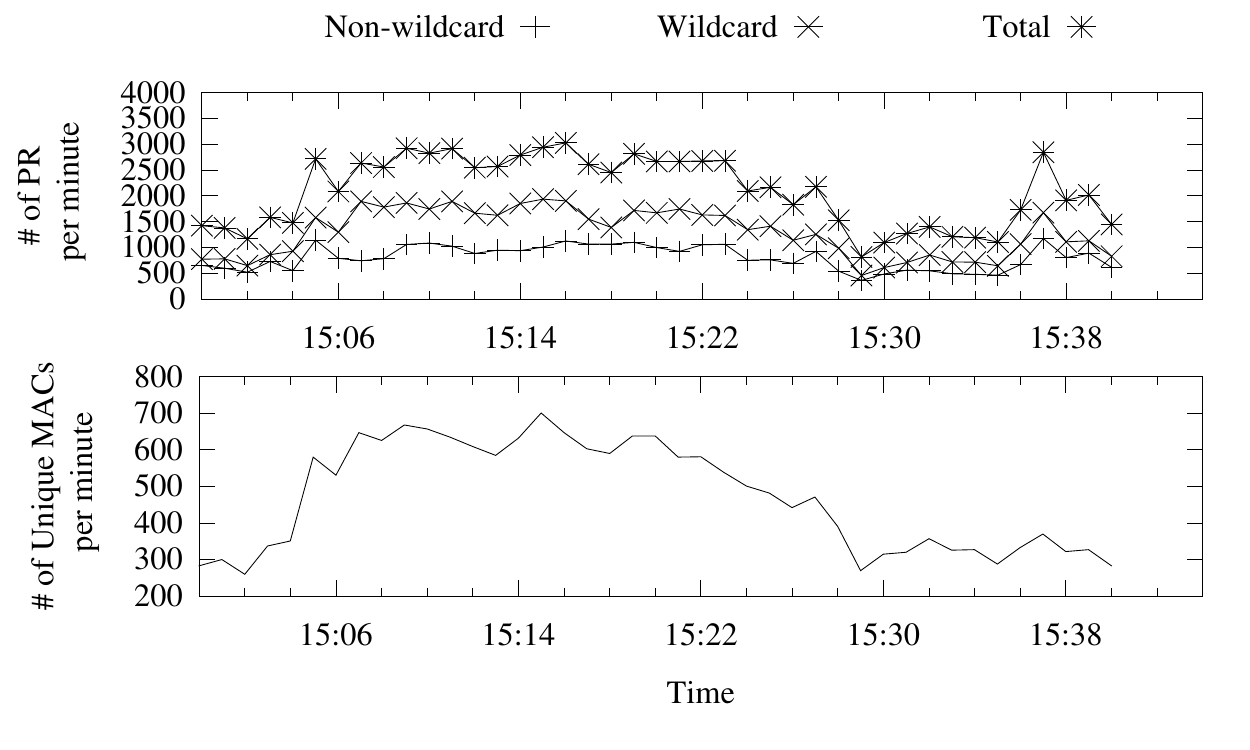}

\caption{Stanford game}

\label{Graph:StanfordTS}

\end{subfigure}

\caption{Minute-wise Avg. \# of Probe Requests and Avg. \# of Unique MAC Addresses on 2.4 GHz Channel 1} 

\label{time_series}

\end{figure}

\begin{figure}
\centering
\includegraphics[scale=0.65]{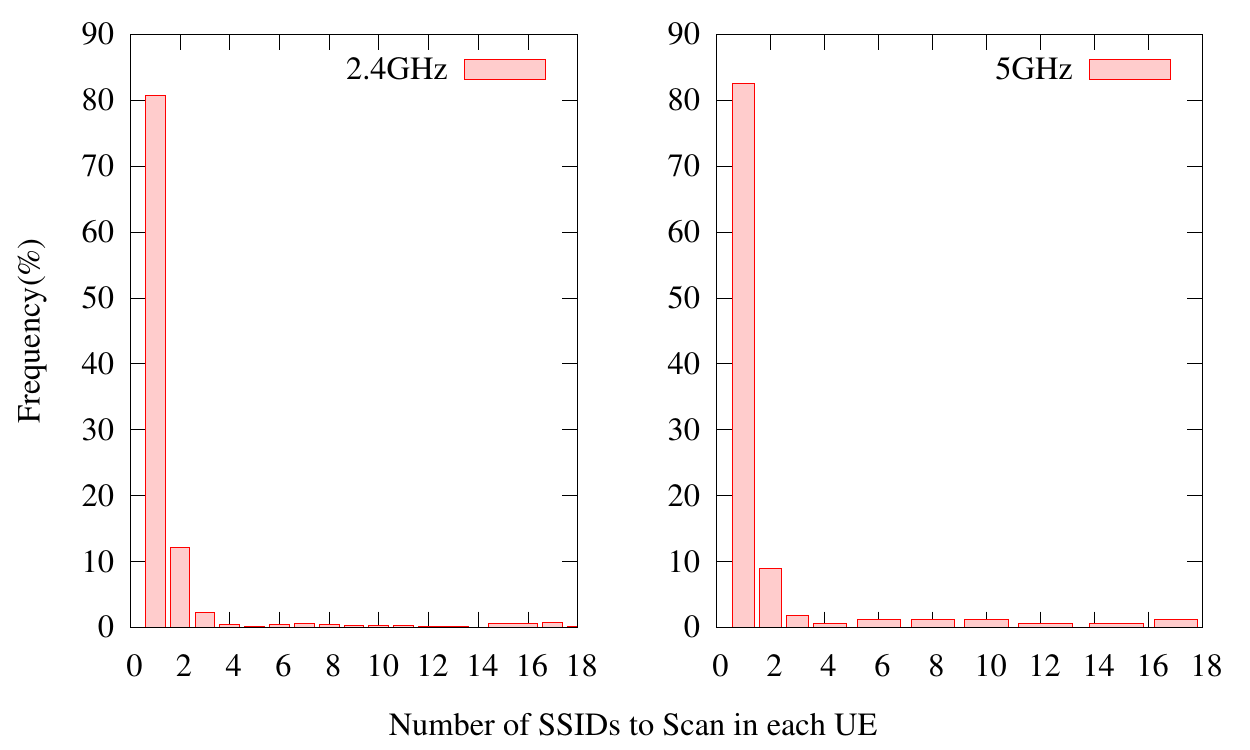}
\caption{\# of SSIDs per Active Scan in Michigan Game on 2.4 GHz (Channel 1) and 5 GHz (Channel 153)}
\label{Graph:SSID_UE}
\end{figure}

Figure \ref{time_series} shows the time series distributions of Probe Requests during the recordings for Channel 1 in the 2.4 GHz as observed for both the Michigan game (Figure \ref{Graph:MichiganTS}) and Stanford game (Figure \ref{Graph:StanfordTS}).  The per-minute average number of Probe Requests is given in terms of the total number of Probe Requests as well as the wildcard (empty SSID) and non-wildcard (known SSID, ex. ND-Secure).  In addition, the number of unique MACs per minute is also plotted over time where a unique MAC is defined as being unique within that particular minute.  We posit that part of the increase in the number of unique MAC addresses at the Stanford game arose from changes introduced by iOS 8 with regards to Probe Request anonymity.  We see that the number of Probe Requests stays relatively constant due to the natural queuing process at the gates which we believe in turn can also serve as an indicator of Probe Request prevalence when individuals are seated at the stadium.  Notable dips can also be observed at the start of the game (7:30 PM for Michigan, 3:30 PM for Stanford).        

\begin{figure}

\centering

\begin{subfigure}[b]{\textwidth}

\includegraphics[scale=0.65]{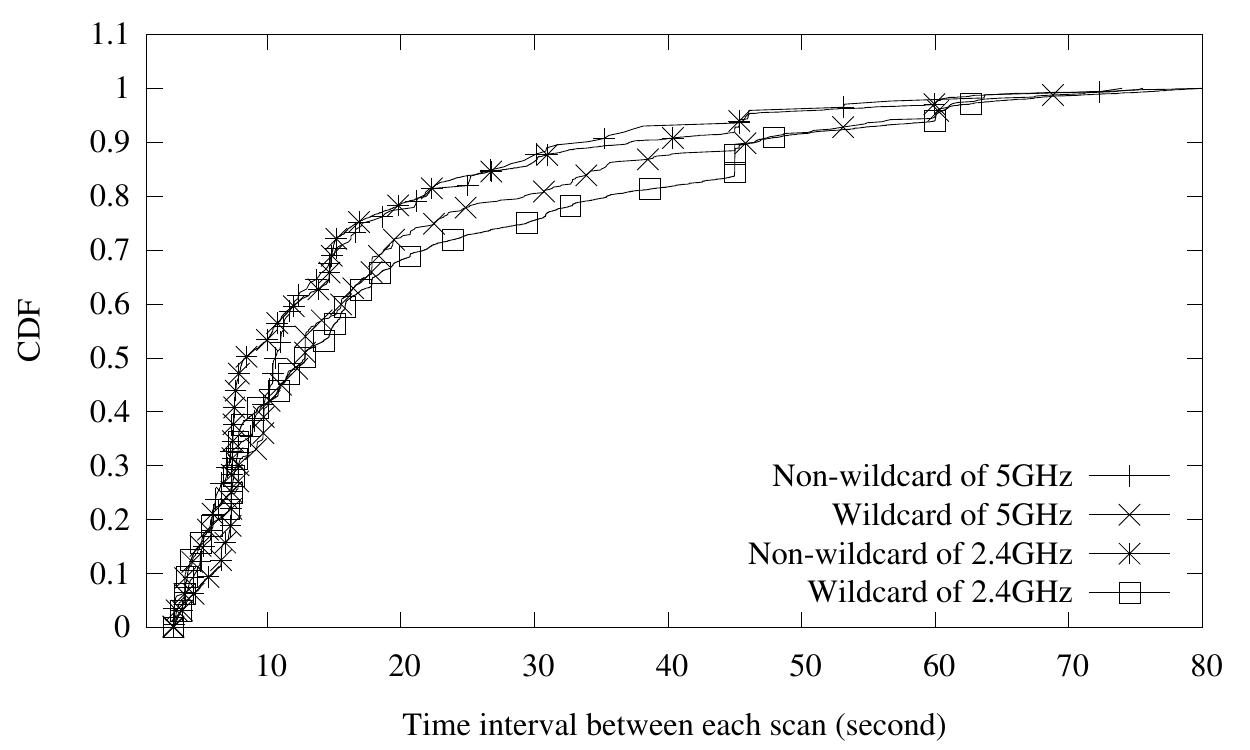}

\end{subfigure}

\begin{subfigure}[b]{\textwidth}

\includegraphics[scale=0.65]{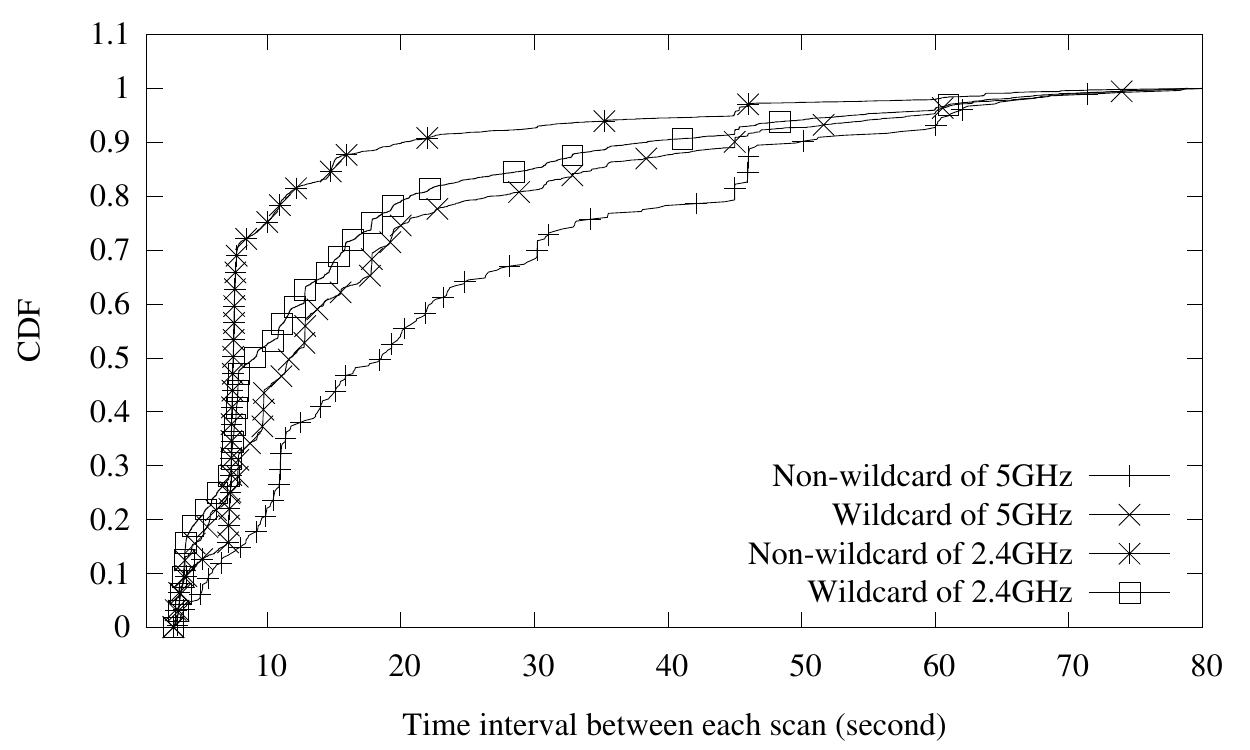}

\end{subfigure}

\caption{Scan Interval}

\label{Graph:ScanInterval}

\end{figure}

Figure \ref{Graph:SSID_UE} continues by plotting the relative probabilities of SSID requests within the observed Probe Request traffic for the Michigan game in order to avoid issues with MAC anonymization.  WiFi scans are divided into sweeps representing a complete, active scan and the number of unique SSIDs requested is mapped to each individual UE.  The leftmost case (1) represents where a mobile node does not request any known SSIDs from its PNL (Preferred Network List).  As can be observed from the graph, most nodes do not make requests for SSIDs from their PNL but rather only make requests for the the empty (unknown) SSID.  Interestingly, several mobile nodes were quite chatty providing extensive PNL requests.  We comment a bit later on scan mechanism and how the length of the PNL can have only a minimal impact on the actual number of resulting scans (ex. some mobile nodes tend to scan for a timed duration rather than simple PNL coverage).         

\begin{figure}
\centering
\includegraphics[scale=0.6]{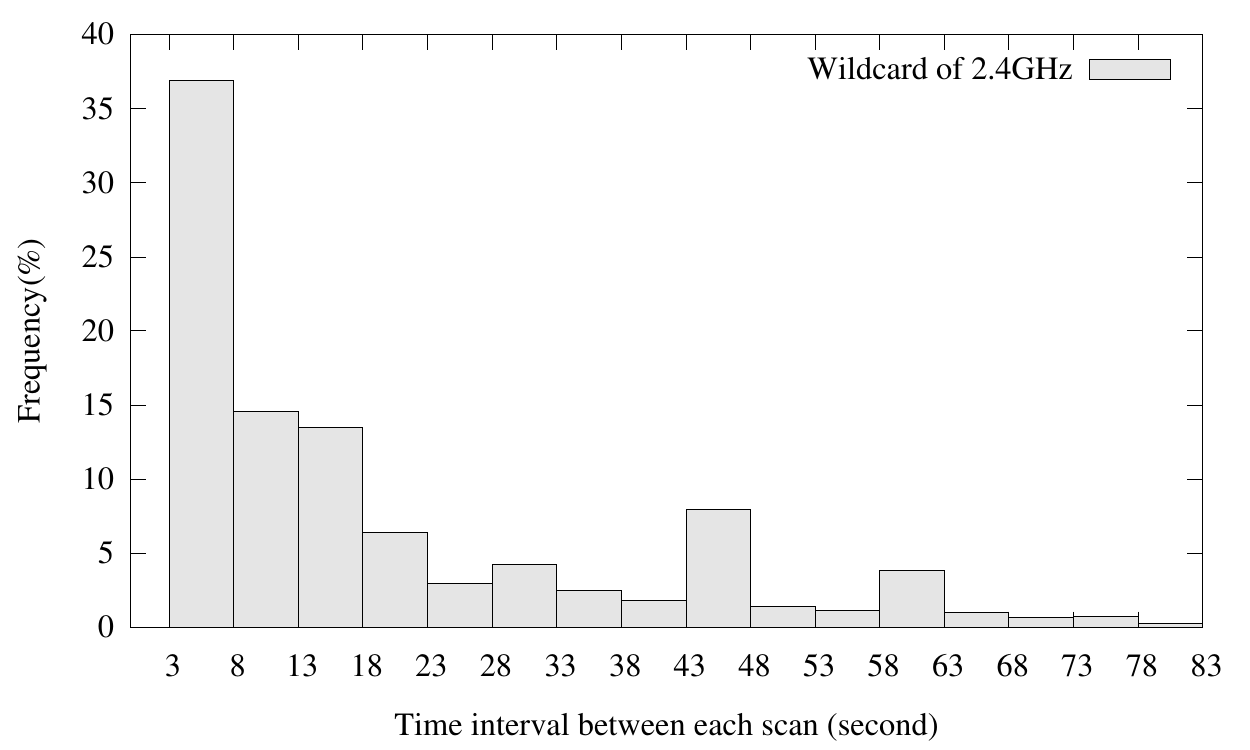}
\caption{Frequency of Scan Interval in 5-sec Window}
\label{Graph:ScanPDFStanford}
\end{figure}

Next, Figure \ref{Graph:ScanInterval} plots a CDF of the inter-scan interval that represents the average wait time between successive scans initiated by mobile nodes for the Michigan and Stanford games across both the 2.4 and 5 GHz spectrums.  Each CDF is also broken out by the requests for empty SSIDs (interval between unknown SSIDs) and known SSIDs via the PNL.  A low-pass filter is applied with a floor of three seconds as observed by the data distributions that means the interval is only counted if the node has at least three seconds of idle time between requests from the same UE with the same SSID.  SSIDs from the PNL may only be counted once for a given UE (ex. BestBuy matching with BestBuy counts and then precludes any subsequent matches in the same scan for that UE).  The interesting result is that many nodes (50\% for Michigan, 70\% for Stanford) scanned quite frequently at 10 second intervals.  Figure \ref{Graph:ScanPDFStanford} breaks out a histogram of scan intervals using 5 second windows for bucketing purposes.  The frequent scan may in part be driven the cellular network on campus being overwhelmed on game day (nearly 150k individuals can be on campus) as well as individuals turning on their phone trying to find Wifi while waiting in line to enter the stadium.      

\begin{figure}

\centering

\includegraphics[scale=0.6]{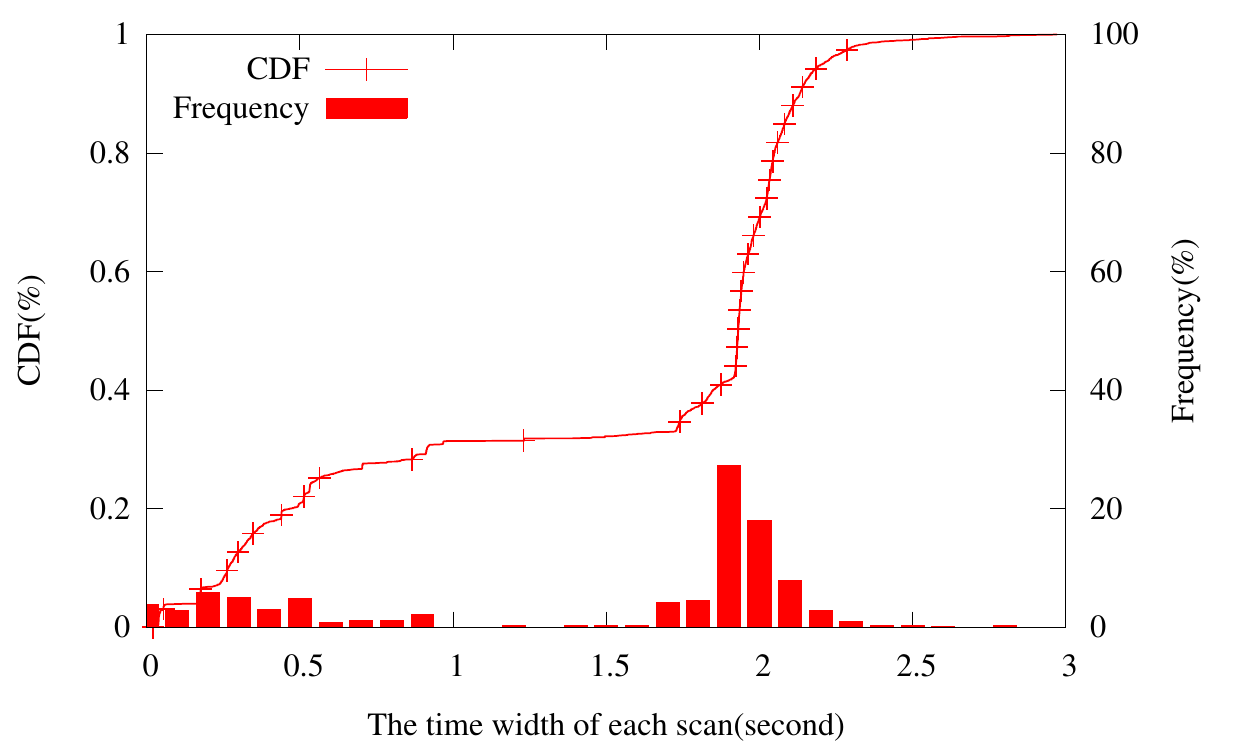}

\caption{Time Width of WiFi Scan}

\label{Graph:TimeSpan}

\end{figure}

Finally, Figure \ref{Graph:TimeSpan} measures the average time width for a particular active scan.  The time width is recorded by measuring the occurrence of the first Probe Request for a UE in Channel 1 followed by the appearance of the last Probe Request for a UE in Channel 11.  Time synchronization is provided by running each of the monitors on the same laptop in monitor mode.  The data for Figure \ref{Graph:TimeSpan} was drawn exclusively from the Stanford game and focused only on the 2.4 GHz band.  Both the CDF and frequency (PDF) are plotted in the figure.  Interestingly, there are two clusters that can largely be attributed to differences between the respective mobile operating systems.  On the left side, Android devices tend to frequently try to cap the maximum scan width ranging typically less than 800 milliseconds.  Alternatively, iOS devices tend to fan out over a wider period of time by scanning for up to 2 seconds at a time.  Critically, Android devices tend to cram as many Probe Requests as possible into a shorter period of time while iOS devices tend to spread out the Probe Requests over time.    

\section{Performance Deterioration}

While the previous section provided insight into WiFi scanning as observed in the ultra-dense stadium environment, we pivot to explore the energy and throughput costs of aggressive WiFi scanning in the controlled laboratory setting. Although it would be ideal to instrument the entirety of the stadium and to provide pervasiveness, the laboratory experiments can shed light on what might occur in the larger scale scenarios. To that end, we conducted a group of small-scale, controlled experiments with three types of handsets: the iPad mini, the Dell Venue 7 tablet, and the Samsung Galaxy S4 smartphone. We are particularly interested in measuring the power cost of an active WiFi scan as well as the throughput impacts associated with WiFi scans. 

For our experiments, we investigated WiFi scanning behaviors for the aforementioned devices by configuring the devices with two different settings while at the same time running \emph{tcpdump} in monitor mode for capturing Probe Requests. The laptops utilized for packet capture were identical to the setups used for packet capture in the stadium environment. Scanning behaviors of the devices are summarized in Table \ref{table:scan}. The distinguishing factor between the two columns refers to whether or not the WiFi settings screen was open (which implies a much more aggressive approach to scan). As indicated by the table, for all three types of devices (WiFi enabled), if the listing of current WiFi is open, the intervals between two consecutive WiFi scans are roughly 10 seconds. Hence, one could largely identify cases from the stadium intervals of 10 seconds as highly likely as being driven by users opening their WiFi settings. Alternatively though more difficult to measure, the lack of cellular connectivity may also have caused more rapid scanning. If WiFi is still on but the user is not in the WiFi settings screen (but not connected to WiFi), the scan interval for different devices varies from 20 seconds to up to 4 minutes. Neither the iPad mini nor the Dell Venue 7 tablet for the experiment had a cellular adapter. For all our experiments presented in this section, we forced the handset to stay in WiFi settings screen since this configuration allows us to better mimic denser environmental scenarios.  

\begin{table}
\centering
\caption{WiFi Scan Interval}
\label{table:scan}
\begin{tabularx}{0.45\textwidth}{|c *{3}{|>{\centering\arraybackslash}X}|}
\hline
\multicolumn{1}{|l|}{} & \multicolumn{2}{c|}{\textbf{In WiFi Settings Screen}} \\ \hline
\textbf{Device Type} & \textbf{Yes} & \textbf{No} \\ \hline
{iPad mini} & 10 s & 20 s to 240 s \\ \hline
{Dell Venue 7} & 10 s & 45 s \\ \hline
{Galaxy S4} & 10 s & 18 to 130 s \\ \hline
\end{tabularx}
\end{table}

\subsection{Energy Impact}

\begin{figure}
\centering 
{\includegraphics[width=3.25in]{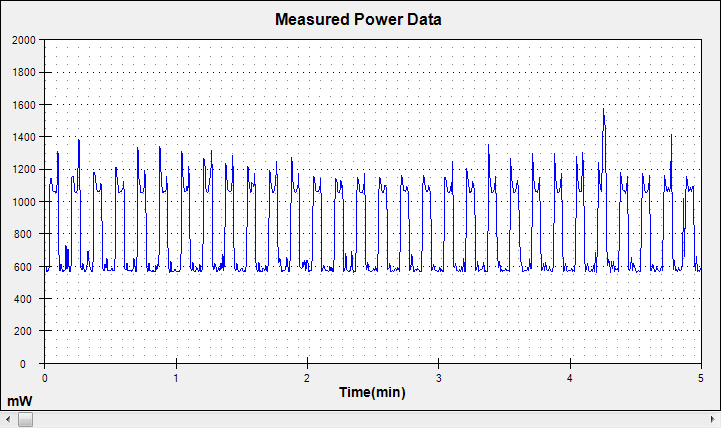}}
\caption{Power Trace of WiFi Scan} 
\label{fig:screenshot}
\end{figure}

For the purpose of evaluating energy cost per WiFi scan, we use the Monsoon power monitor and its PowerTool software.  We instrument the Galaxy S4 smartphone as it is the only device capable possessing a removable battery that the Monsoon can accurately determine power consumption. Power for the phone is supplied by the Monsoon power monitor with the energy consumption recorded at a sampling rate of 5 KHz. The phone is evaluated in two different settings, \emph{Baseline} where WiFi is off but the screen remains on and \emph{Scanning} where the phone stays in the WiFi settings screen but remains unaffiliated with any WiFi AP. The PNL for the smartphone consisted of two SSIDs though our experiments (not shown) found that the PNL would need to grow significantly (12+) in order to affect scan width.  Power monitoring was run for an extended period of time (5 minutes) and as shown in Figure \ref{fig:screenshot}, considerable spikes can be observed whenever an active scan is initiated. The average and standard deviation of power consumption with each setting are given in Table \ref{table:power}. 

After recording the average power consumption for both the Baseline and the Scanning settings, the energy cost of an active WiFi scan can be approximated by calculating the delta between the consumption values of these two settings over the entirety of the monitoring period, yielding 5467.76 uAh for the 5-minute time window.  Notably, this represents a 44.3\% increase over the Baseline consumption despite the fact that the screen is on for both cases. We can further infer the power cost per scan since Table \ref{table:scan} has indicated WiFi scan is invoked every 10 seconds if WiFi setting screen is active. This type of behavior is also verified by Figure \ref{fig:screenshot}, which is the power monitoring graph generated by Monsoon for a 5-min experiment using the Scanning setting. It is observed from Figure \ref{fig:screenshot} there are consistent power spikes caused by WiFi scanning every 10 seconds, giving a total number of thirty scans across the 5-minute monitoring period. While somewhat crude in its approximation, each WiFi scan in an ideal scenario (no background traffic, no DCF issues) consumes roughly 182.26 uAh.  An individual waiting in line for 10 minutes without WiFi while aggressively scanning could end up consuming nearly 10936 uAh extra energy, effectively 0.4\% of a fully charged battery (2600 mAh). Now, envision this scenario though continuing throughout a game (3 hours) and roughly 7\% of the energy capacity of the phone is wasted with unrequited WiFi Probe Requests. Consider this same scenario applied to every smartphone within the stadium (80K devices) and suddenly the wasted energy appears a bit more significant.      
  
\begin{table}[t]
\centering
\caption{Energy Cost: Baseline vs. Scanning}
\label{table:power}
\begin{tabularx}{0.45\textwidth}{|c *{3}{|>{\centering\arraybackslash}X}|}
\hline
\multicolumn{1}{|c|}{} & \multicolumn{2}{c|}{\textbf{Power Consumption (uAh)}} \\ \hline
\textbf{Metrics} & \textbf{Baseline} & \textbf{Scanning}  \\ \hline
{Average} & 12333.42 & 17801.18   \\ \hline
{Std Dev} & 7.18 & 60.08  \\ \hline
\end{tabularx}
\end{table}

\subsection{Throughput Loss}

While the energy losses may be tolerable (though still wasteful), aggressive WiFi scanning also has significant impacts on networking performance given that it constantly introduces bandwidth overhead to wireless channels by virtue of the frequent Probe Requests. For the purpose of measuring potential network performance degradation introduced by Probe Requests, we designed a small-scale experiment using a UDP client / server arrangement with UDP throughput as the defined performance metric. The components of this experiment are described as follows: 

\emph{UDP client and server:} We instrumented a Dell EliteBook 8560 with a 802.11n dual-band network interface as the client and a HP 3450 laptop with Kali Linux installed as the server. The rational for choosing UDP lies in the fact with UDP we are able to control the packet size and the speed of sending packets. The client was implemented in C\# for sending packets with specified length and speed. The server was written in Java to receive packets and record the arriving time and packet size.
	
\emph{Handsets:} We used one iPad mini, one Dell Venue 7 tablet, and three Samsung Galaxy S4 smartphones as our handsets. All devices have dual-band capability and as a result send Probe Requests across both the 2.4 and 5 GHz bands. For each device, we de-associated the device from any known WiFi network and create a PNL consisting of two hidden SSIDs.  Screens were kept on and the devices were kept in the WiFi scan screen.  
	
\emph{Wireless router:} A Netgear AC1900 router (dual-band, 802.11ac capable) was used to set up the WLAN for our experiment. The router provided up to 600 Mbps and 1300 Mbps WiFi down-link speeds on the 2.4 and 5 GHz bands respectively.  
	
\emph{Connection:} The client was the only node associated with the SSID for the 5 GHz band of the router while the server was directly connected to the router via Gigabit Ethernet. A full 40 MHz of spectrum (full 802.11n speeds) was granted to the router with validation for the lack of interference on the 5 GHz channel.  Experiments were conducted in a basement with minimal interference from other devices.  The distance between the client and the server was roughly four meters and all handsets were positioned between the client and the server.   

The client was tuned in order to determine the maximum lossless send rate between the client and the server. Communications were unidirectional going only from the client to the server. Client performance topped out at roughly 130 Mb/s without any background traffic (see Table \ref{table:tp}). Each experiment setting was repeated five times with a typical test duration of five minutes. Once the baseline was established, the experiments were repeated with varying numbers of UEs ranging from 1 UE (iPad Mini) to the case of the entire bank of devices (in the order by which devices were introduced: iPad Mini, Dell Venue 7 tablet, 3x Samsung Galaxy S4 smartphones). 

Notably, WiFi performance decreases dramatically once the first few devices are introduced. With the addition of the first Galaxy S4 smartphone, performance has already decreased from the peak of over 130 Mb/s down to 97 Mb/s. While degradation of performance is not entirely unexpected with WiFi, the fact that this performance decrease comes by virtue of `useless' Probe Requests is especially problematic. For stadium environments where not all nodes are directly affiliated with the preferred venue WiFi, there may be notable speed reductions by virtue of unaffiliated mobile nodes still chirping for WiFi. Furthermore, our limited lab experiments were actually quite benign representing at best 20+ unaffiliated clients (5 UEs probing at four times the effective rate, 40s vs. 10s when in the WiFi settings screen). In the stadium case where hundreds of nodes may be present, unaffiliated nodes could have significant performance issues even in the `better' 5 GHz bands. We leave out the 2.4 GHz results due to space constraints which fare much worse due to the lack of orthogonality in said channels.            

\begin{table}[t]
\centering
\caption{Throughput Reduction}
\label{table:tp}
\begin{tabularx}{0.45\textwidth}{|c *{4}{|>{\centering\arraybackslash}X}|}
\hline
\textbf{} & \multicolumn{2}{c|}{\textbf{Server Throughput (Mb/s)}} \\ \hline
\multicolumn{1}{|l|}{\textbf{Number of UEs}} & \textbf{Average} & \textbf{Std Dev} \\ \hline
\textbf{0} & 132.70 & 2.73 \\ \hline
\textbf{1} & 123.18 & 2.03 \\ \hline
\textbf{2} & 113.27 & 14.23 \\ \hline
\textbf{3} & 97.67 & 10.46 \\ \hline
\textbf{4} & 83.06 & 20.36 \\ \hline
\textbf{5} & 67.56 & 18.63 \\ \hline
\end{tabularx}
\end{table}

\section{Discussion}

The issue of how to solve the dilemma of aggressive WiFi speaks to the complexities of the wireless industry.  On one hand, the solution would appear to be fairly trivial:  slow down the WiFi scanning rate and scan only on a screen activation or something similar.  The reality though is decidedly more complicated as applications largely do not wait to send data until there is WiFi available and the delay before locating WiFi at home could be considerable.  Critically, the vast majority of a user experience tends to be dominated by the simple cases, ex. only a few devices and well-known SSIDs.  In those cases, the first scan tends to be successful and faster scanning means faster hopping on WiFi.  

Furthermore, the vendors at first glance most impacted by aggressive Probe Requests tend to be the WiFi infrastructure vendors who have little to no control over the mobile devices.  After all, WiFi exists in the unlicensed bands which means that for all practical purposes, the equipment infrastructure vendors must simply endure.  Recent discussions with the 802.11ax standard have noted that indeed, aggressive Probe Requests do create sizable issues in ultra-dense venues \cite{Cisco:Athens:80211ax}.  The 802.11hew also seeks to further address issues in the ultra-dense cases though from largely a management perspective.  Handset and OS vendors are only marginally by claims of reduced throughput as the vast majority of throughput is OK (the typical case).  However, we would argue that the energy cost of being aggressive is distinctly non-trivial and moreover, that energy cost burns worst when most users tend to suffering energy issues acutely (ex. where a user is unlikely to be able to charge).         

\section{Conclusions and Future Work}

In conclusion, we believe that aggressive WiFi scanning has significant side effects on nominal wireless network users, both with respect to energy and throughput.  Moreover, we believe that our stadium analyses show that not only are aggressive Probe Requests wasteful in the ultra-dense case, the aggressive nature tends to exacerbate the energy costs draining the energy of the entire local mobile device population significantly faster.  Interesting future work is needed to explore how one can bridge the conflicting goals of rapid WiFi detection with the cost of wasted WiFi scans. To help expedite such efforts, we intend to release the recorded gate data as a public dataset after anonymization.  The dataset will be contributed to CRAWDAD and made directly available via our website.

%
\bibliographystyle{abbrv}
\bibliography{bib}  

\begin{thebibliography}{1}

\bibitem{ANDSF}
Cellular-wi-fi integration.
\newblock Technical report, InterDigital, 2012.

\bibitem{HS20}
The future of hotspots: Making wi-fi as secure and easy to use as cellular.
\newblock Technical report, Cisco, 2012.

\bibitem{Andrews:ChallengeCell}
J.~G. Andrews, S.~Buzzi, W.~Choi, S.~V. Hanly, A.~E. Lozano, A.~C.~K. Soong,
  and J.~C. Zhang.
\newblock What will 5g be?
\newblock {\em CoRR}, abs/1405.2957, 2014.

\bibitem{Gupta:SECON07:WifiPhone}
A.~Gupta and P.~Mohapatra.
\newblock Energy consumption and conservation in wifi based phones: A
  measurement-based study.
\newblock In {\em Sensor, Mesh and Ad Hoc Communications and Networks, 2007.
  SECON '07. 4th Annual IEEE Communications Society Conference on}, pages
  122--131, June 2007.

\bibitem{Cisco:Athens:80211ax}
B.~Hart, M.~Swartz, J.~Suhr, and M.~Silverman.
\newblock Stadium measurements.
\newblock 802.11ax Working Group (Doc 1223), September 2014.

\bibitem{Raghavendra:TMC10:UnwantedLink}
R.~Raghavendra, E.~Belding, K.~Papagiannaki, and K.~Almeroth.
\newblock Unwanted link layer traffic in large ieee 802.11 wireless networks.
\newblock {\em Mobile Computing, IEEE Transactions on}, 9(9):1212--1225, Sept
  2010.

\bibitem{Rayanchu:NSDI12:WiFiNet}
S.~Rayanchu, A.~Patro, and S.~Banerjee.
\newblock Catching whales and minnows using wifinet: Deconstructing non-wifi
  interference using wifi hardware.
\newblock In {\em Presented as part of the 9th USENIX Symposium on Networked
  Systems Design and Implementation (NSDI 12)}, pages 57--70, San Jose, CA,
  2012. USENIX.

\bibitem{Teng:INFOCOM09:DScan}
J.~Teng, C.~Xu, W.~Jia, and D.~Xuan.
\newblock D-scan: Enabling fast and smooth handoffs in ap-dense 802.11 wireless
  networks.
\newblock In {\em INFOCOM 2009, IEEE}, pages 2616--2620, April 2009.

\bibitem{Yeo:WiSe04:LANMonitor}
J.~Yeo, M.~Youssef, and A.~Agrawala.
\newblock A framework for wireless lan monitoring and its applications.
\newblock In {\em Proceedings of the 3rd ACM Workshop on Wireless Security},
  WiSe '04, pages 70--79, New York, NY, USA, 2004. ACM.

\end{thebibliography}

%
%

\end{document}